\documentclass{article}

\usepackage{arxiv}

\usepackage[utf8]{inputenc} 
\usepackage[T1]{fontenc}    
\usepackage{hyperref}       
\usepackage{url}            
\usepackage{booktabs}       
\usepackage{nicefrac}       
\usepackage{microtype}      
\usepackage{graphicx}
\usepackage{doi}

\def\bSig\mathbf{\Sigma}

\usepackage[figuresright]{rotating}
\usepackage{amsmath,amsfonts,bm}
\usepackage{multirow}
\usepackage{natbib,psfrag}
\usepackage{comment}
\usepackage{fancyhdr,amssymb}
\usepackage[titletoc]{appendix}
\usepackage{booktabs}
\usepackage{threeparttable}
\usepackage{bbold}
\usepackage{mathtools}

\newtheorem{theorem}{Theorem}

\title{Interim Monitoring in Sequential Multiple Assignment Randomized Trials}

\author{ Liwen Wu\thanks{Department of Biostatistics, University of Pittsburgh, Pittsburgh, PA 15261. Email: liw88@pitt.edu} \\
	\And
	Junyao Wang \\
	\And
	Abdus S. Wahed \\

}

\date{}

\renewcommand{\shorttitle}{\textit{IM-SMART}}

\hypersetup{
pdftitle={Interim Monitoring in Sequential Multiple Assignment Randomized Trials},
pdfsubject={Statistics, Methodology},
pdfauthor={Liwen Wu, Junyao Wang, Abdus S. Wahed},
pdfkeywords={Adaptive treatment strategy; Dynamic treatment regime; Group sequential analysis; Interim monitoring; Sequential multiple assignment randomized trial},
}

\begin{document}
\maketitle

\begin{abstract}
A sequential multiple assignment randomized trial (SMART) facilitates comparison of multiple adaptive treatment strategies (ATSs) simultaneously. Previous studies have established a framework to test the homogeneity of multiple ATSs by a global Wald test through inverse probability weighting. SMARTs are generally lengthier than classical clinical trials due to the sequential nature of treatment randomization in multiple stages. Thus, it would be beneficial to add interim analyses allowing for early stop if overwhelming efficacy is observed. We introduce group sequential methods to SMARTs to facilitate interim monitoring based on multivariate chi-square distribution. Simulation study demonstrates that the proposed interim monitoring in SMART (IM-SMART) maintains the desired type I error and power with reduced expected sample size compared to the classical SMART. Lastly, we illustrate our method by re-analyzing a SMART assessing the effects of cognitive behavioral and physical therapies in patients with knee osteoarthritis and comorbid subsyndromal depressive symptoms.
\end{abstract}


\section{Introduction}
\label{s:intro}
Adaptive treatment strategies (ATSs) are sequences of treatments personalized to patients given their own treatment history. Such strategies are popular in the management of chronic diseases, such as mental health disorders, where the treatment courses usually involve multiple stages due to the dynamic nature of disease progression  \citep{lavori2000flexible, lavori2004dynamic, murphy2005experimental, kidwell2014smart}. A sequential multiple assignment randomized trial (SMART) is a systematic and efficient vehicle for comparing multiple ATSs. In a SMART, after initial randomization and evaluation of an intermediate outcome, patients may be re-randomized to future treatments at the following stages. Several well-known applications of SMART include the CATIE trial for Schizophrenia \citep{schneider2003clinical}, the STAR*D for major depression disorder \citep{rush2004sequenced}, the CCNIA trial in children with autism \citep{kasari2014communication}, the ADEPT trial for mood disorders \citep{kilbourne2014protocol}, and many recent oncology studies \citep{kidwell2014smart}. 

The challenge of evaluating ATSs from a SMART lies in taking into account the adaptive and sequential nature of treatment assignment. For continuous outcomes, \cite{murphy2005experimental} first presented the strategy mean estimators based on the inverse probability weighting (IPW) method and proposed a corresponding sample size formula to test two ATSs from a SMART design. \cite{ dawson2010efficient, dawson2010sample} proposed a semi-parametric approach to estimate the ATS effects from SMART and used `variance influence factor' to account for the loss of precision due to sequential randomization. \cite{ko2012up} demonstrated that the inverse probability weight-normalized (IPWN) estimators are more efficient than the IPW estimators. They also provided a Wald test to compare multiple ATSs simultaneously. Based on the same testing procedure, \cite{ogbagaber2016design} further developed the power and sample size formulas for testing multiple ATSs. Methods for binary and survival outcomes have also been established in the past decade. \citep{kidwell2013weighted, kidwell2018design}.

A caveat of SMARTs is that, given the sequential nature of treatment decisions, many trials require more time to be completed than that for classical randomized trials. Compared to single-stage randomized controlled trials, patients are followed up through multiple stages in SMARTs, and the total length of study duration depends on the number of stages, definition of intermediate outcomes, as well as other clinical and ethical considerations that are equally important in traditional designs \citep{kidwell2014smart}. One way to improve trial efficiency is to introduce interim analyses where early stops for efficacy and/or futility could be made by repeatedly testing the null hypothesis as data accumulates \citep{armitage1960sequential}. A large family of methods have been developed over the years to serve this purpose \citep[among others]{pocock1977group, o1979multiple, gordon1983discrete, wang1987approximately}, where the multiple looks into the data are facilitated by adjusting the nominal type I error at each interim look. The critical values are obtained from the joint distribution of the test statistics from repeated assessments.

Currently, there is no interim monitoring method in SMARTs for testing multiple ATSs simultaneously. Some statistical methods exist for interim analyses in multi-arm multi-stage (MAMS) clinical trials, a design similar to SMART but without the sequential randomization feature. \cite{wason2012optimal} proposed a `simultaneous' single step stopping rule with an error spending function, while \cite{urach2016multi} optimized the multiplicity problem by incorporating sequential rejective tests. However, both of these methods are based on pair-wise comparisons between different treatment strategies, or between different active treatments and a common control arm, instead of a global test of strategy effects. In the SMART setting, \cite{ertefaie2016identifying} proposed a promising re-sampling based method to identify the subset of best ATSs while adjusting for the multiplicity as well as the correlation between strategies. However, it is not clear how this method can be generalized to repeated testing. Bayesian method for interim monitoring and adaptation has been proposed in a relevant design called small n Sequential Multiple Assignment Randomized Trial (snSMART) that compares different treatments in a multi-stage trial with a limited sample size. The objective of such a design is, however, inherently different from SMART because it compares individual treatments rather than treatment sequences \citep{wei2018bayesian, chao2020bayesian}. Thus, development of sequential testing procedures for global test of ATSs in SMARTs is warranted from pragmatic considerations. 

In this study, we introduce the traditional group sequential methods for clinical trials to SMARTs. In Section \ref{s:method}, we describe the classical SMART design and the corresponding statistical inference. In Section \ref{GS-SMART}, we introduce a method to perform interim monitoring in SMART (IM-SMART). The results from a simulation study are presented in Section \ref{s:simulation}. In Section \ref{s:rapid}, we apply our proposed method to a dataset from the RAPID study assessing the effects of cognitive behavioral and physical therapies in patients with knee osteoarthritis and comorbid subsyndromal depressive symptoms to illustrate the benefits of including interim analysis in SMARTs. We conclude the paper with a discussion in Section \ref{s:discuss}. 
 
\section{Classical SMART Design and Inference}
\label{s:method}

\subsection{Notation}
\label{subs:notation}
SMART designs vary depending on the intermediate response, the number of stages, and the number of treatments available at each stage \citep{ogbagaber2016design}. For clarity in describing our method, we use a SMART design similar to the EXTEND trial \citep{oslin2005managing, lei2012smart}, where patients are initially randomized to two treatments $A_j, j=1,2$, and then to two maintenance treatments $B_k, k=1,2$, if they responded, or to two alternative treatments $C_l, l=1,2$, if they failed to respond to the initial treatments (Figure 1). The objective is to estimate and compare overall strategy effects. In this trial, we have the opportunity to compare 8 different treatment strategies $A_jB_kC_l, j,k,l=1,2$, where $A_jB_kC_l$ is defined as `treat with initial treatment $A_j$, then $B_k$ if respond to $A_j$, or $C_l$ if do not respond to $A_j$.' 
\begin{figure}
 \centerline{\includegraphics[width=5.25in]{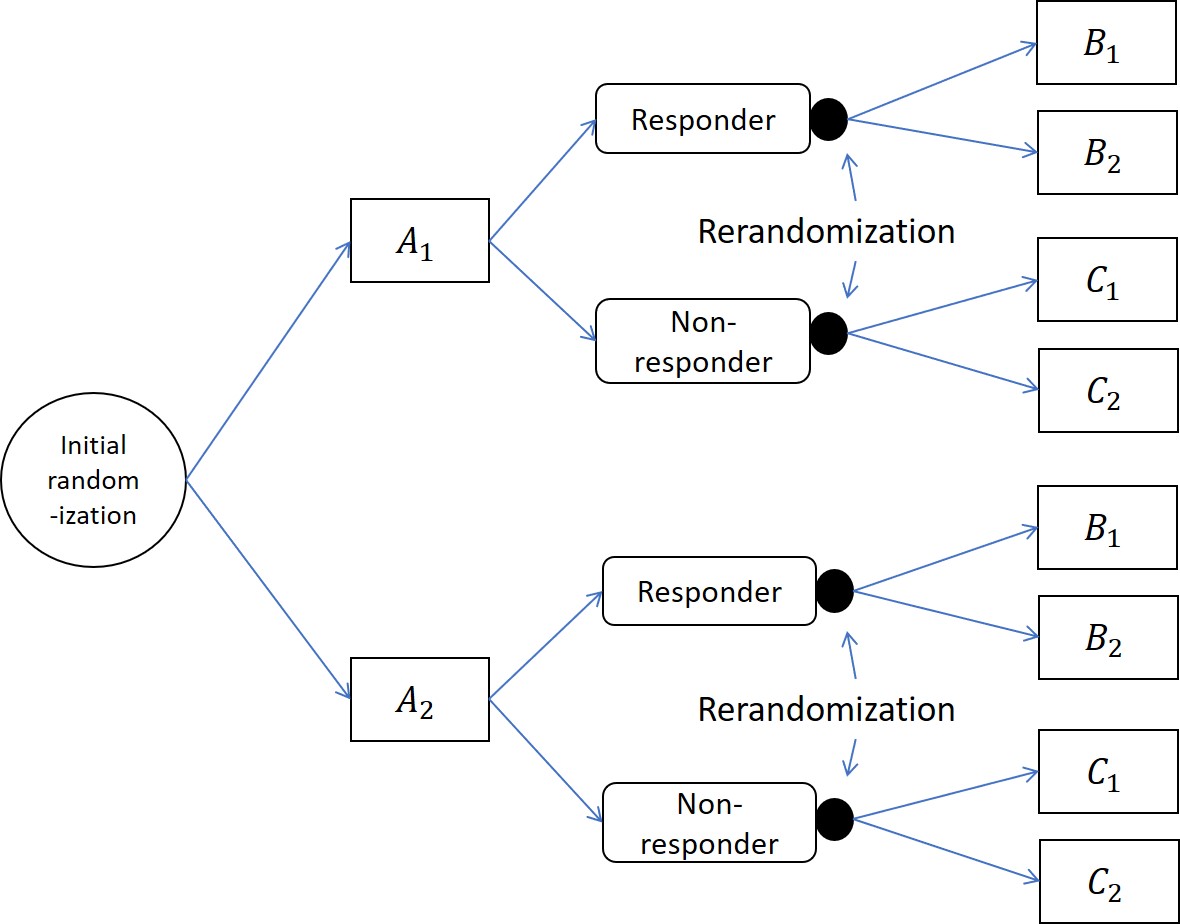}}
\caption{Example of SMART design similar to the EXTEND trial \citep{lei2012smart}. There are two initial treatments $A_j, j=1,2$, two maintenance treatments $B_k, k=1,2$ for patients who respond to their initial treatments, and two alternative treatments $C_l, l=1,2$ for non-responders. There are in total 8 different ATSs embeded: $A_jB_kC_l, j,k,l=1,2$.}
\label{f:design}
\end{figure}

To facilitate the comparison of ATSs, we cast the problem in a counterfactual framework. Define $Y$ as the outcome of interest. Without loss of generality, we will treat $Y$ as a continuous variable in this paper. Let us define $Y_i(A_jB_k)$ to be the potential outcome if patient $i$ received initial treatment $A_j$, responded, and received maintenance treatment $B_k$. Similarly, we define potential outcome $Y_i(A_jC_l)$ for patient $i$ who received initial treatment $A_j$, did not respond, and received alternative treatment $C_l$. Let $R_{ji}$ denotes the counterfactual response status for patient $i$ under treatment $j$ (ie. $R_{ji}=1$ if the patient responded to the initial treatment $A_j$, and 0 otherwise). The counterfactual outcome $Y$ for patient $i$ following strategy $A_jB_kC_l$ can be written as
\begin{equation}
Y_i(A_jB_kC_l) = R_{ji}Y_i(A_jB_k)+(1-R_{ji})Y_i(A_jC_l), \; j,k,l=1,2.
\end{equation}

Denote $\pi_j$ as the probability of response to the initial treatment $A_j$.  The effect of ATS $A_jB_kC_l$ is then
\begin{equation}
\mu_{jkl}=E(Y(A_jB_kC_l))=\pi_j \mu_{A_jB_k}+(1-\pi_j) \mu_{A_jC_l}, \; j,k,l=1,2,
\end{equation}
\noindent where $\mu_{A_jB_k}=E(Y(A_jB_k))$ and $\mu_{A_jC_l}=E(Y(A_jC_l))$. Inference procedure for these strategies will be described in the following subsections.

\subsection{Observed data and estimators}
\label{s: estimator}
The observed treatment assignment and response for patient $i$ can be recorded as a vector $ \{ I_i(A_j), R_i, R_iI_i(B_k), (1-R_i)I_i(C_l), Y_i, j,k,l=1,2\} $, where each treatment indicator $I_i(\cdot)$ equals 1 if it corresponds to the actual treatments that patient $i$ was randomized to, and 0, otherwise; $R_i$ is the $i^{th}$ patient's response to their initial treatment, $R_i=1$, if the patient is a responder, and 0, otherwise. $Y_i$ denotes the observed outcome for patient $i$ at the end of stage 2. As described in \cite{ko2012up} and \cite{ogbagaber2016design}, the treatment effect of strategy $A_jB_kC_l$, $\mu_{jkl}$, can be estimated by inverse probability weighting using the estimator
\begin{equation}
\label{IPWN}
\hat{\mu}_{jkl} = \frac{\sum^{n}_{i=1} I_i(A_j) \{ \frac{R_iI_i(B_k)}{p_k} + \frac{(1-R_i)I_i(C_l)}{q_l}  \} Y_i}{\sum^{n}_{i=1} I_i(A_j) \{ \frac{R_iI_i(B_k)}{p_k} + \frac{(1-R_i)I_i(C_l)}{q_l}  \}},
\end{equation}

\noindent where $p_k$ is the known randomization probability for $B_k$, and $q_l$ is that for $C_l$. Under the usual assumptions of causal inference, more specifically, consistency, sequential randomization, and positivity, this estimator is shown to be consistent and can be twice as much efficient as a similar estimator where the denominator is just the sample size $\sum^n_{i=1} I_i(A_j)$ \citep{ko2012up}. \cite{ogbagaber2016design} showed that, following the theory of M-estimation, the asymptotic variance of the above estimator is given by
\begin{equation}
\begin{split}
\label{asymvar}
var(\hat{\mu}_{jkl}) = \frac{\sigma^2_{ijk}}{n}
=\frac{\kappa_j}{n} \Bigg[ \frac{\pi_j}{p_k} & \left\{ \sigma^2_{A_jB_k}+(1-\pi_j)^2 (\mu_{A_jB_k}-\mu_{A_jC_l})^2 \right\} \\ & + \frac{1-\pi_j}{q_l} \left\{ \sigma^2_{A_jC_l}+\pi_j^2 (\mu_{A_jB_k}-\mu_{A_jC_l})^2 \right\} \Bigg],
\end{split}
\end{equation}

\noindent where $\kappa_j$ is the inverse of randomization probability to $A_j$,  $\sigma^2_{A_jB_k}=var(Y(A_jB_k))$ and $\sigma^2_{A_jC_l}=var(Y(A_jC_l))$. The asymptotic covariance for two strategies sharing the same initial treatment is, for example, 
\begin{equation}
\label{asymcov}
cov(\hat{\mu}_{111}, \hat{\mu}_{112}) = \frac{\sigma_{111,112}}{n} =
\frac{\kappa_1}{n} \frac{\pi_1}{p_1} \left\{ \sigma^2_{A_1B_1} + (1-\pi_1)^2(\mu_{A_1B_1}-\mu_{A_1C_1})(\mu_{A_1B_1}-\mu_{A_1C_2})  \right\}.
\end{equation}

\noindent The variance and covariance in Equations (\ref{asymvar}) and (\ref{asymcov}) can be estimated by first estimating $\pi_j$,  $\mu_{A_jB_k}$, $\mu_{A_jC_l}$, variances $\sigma^2_{A_jB_k}$, $\sigma^2_{A_jC_l}$ and the randomization probabilities by their empirical counterparts, and plugging those into the above formula \citep{ogbagaber2016design}. Alternatively, following \cite{ko2012up}, the robust estimators for the variance in Equation (\ref{asymvar}) and the covariance in Equation (\ref{asymcov}) can be expressed as
\begin{equation}
\label{var}
    \hat{var}(\hat{\mu}_{jkl})=\frac{\hat{\sigma}^2_{ijk}}{n} =\frac{1}{n_j(n_j-1)}\sum^{n}_{i=1} I_i(A_j) 
    \left\{ \Big(\frac{R_iI_i(B_k)}{p_k} + \frac{(1-R_i)I_i(C_l)}{q_l}  \Big) (Y_i-\hat{\mu}_{jkl})\right\}^2,
\end{equation}
and
\begin{equation}
\begin{split}
\label{cov}
\hat{cov}(\hat{\mu}_{jkl},  \hat{\mu}_{jk'l'}) = \frac{\hat{\sigma}_{jkl,jk'l'}}{n}
=\frac{1}{n_j^2}\sum^{n}_{i=1}  I_i(A_j) & \bigg\{ \big( \frac{R_iI_i(B_k)}{p_k} + \frac{(1-R_i)I_i(C_l)}{q_l}  \Big)  (Y_i-\hat{\mu}_{jkl})\\
&\times \Big( \frac{R_iI_i(B_{k'})}{p_{k'}} + \frac{(1-R_i)I_i(C_{l'})}{q_{l'}}  \Big) (Y_i-\hat{\mu}_{jk'l'})  \bigg\},
\end{split}
\end{equation}

\noindent respectively, where $n_j=n/\hat{\kappa}_j=\sum^n_{i=1} I_i(A_j)$. Notice that, by design, we have  $\hat{cov}(\hat{\mu}_{1kl}, \hat{\mu}_{2k'l'})=0$ for any $k,l,k',l' = 1,2$, since data from patients randomized to different initial treatments are independent from each other. It is also easy to see that $\hat{cov}(\hat{\mu}_{j1l}, \hat{\mu}_{j2l'})=0$ for any $j,l,l' = 1,2$ and $l \ne l'$ due to the sequential randomization assumption \citep{ko2012up}.

\subsection{Hypotheses testing}
\label{inference}

The global null hypothesis to test the homogeneity of all treatment strategies can be written in the form of a linear combination of strategy means, $H_0:\bm{C\mu}={\bm{0}}$, where $\bm{\mu} = [ \mu_{111}, \mu_{112}, \mu_{121}, \mu_{122}, \mu_{211}, \mu_{212}, \mu_{221}, \mu_{222}   ]^T$, and $\bm{C}$ is a contrast matrix. An example could be $\bm{C}= [ \mathbb{1}_7, -1*\bm{I}_{7} ]$, where $\mathbb{1}_7$ is a $7 \times 1$ vector of ones, and $\bm{I}_{7}$ is a $7 \times 7$ identity matrix. 

To test the null hypothesis, we define the following Wald-type test statistic
\begin{equation}
\label{statistics}
    T=n\bm{\hat{\mu}}^T\bm{C}^T [\bm{C\hat{\Sigma}C}^T]^{g} \bm{C \hat{\mu}},
\end{equation}

\noindent where $\bm{\hat{\mu}}$ is the estimated strategy mean vector, and $\bm{\hat{\Sigma}}$ is an estimate of the variance-covariance matrix $\bm{\Sigma}$. More specifically, $\bm{\hat{\Sigma}}=block \; diag (\bm{\hat{\Sigma}}_1, \bm{\hat{\Sigma}}_2)$, where
$$\bm{\hat{\Sigma}}_j=\begin{bmatrix}
\hat{\sigma}^2_{j11}&\hat{\sigma}_{j11,j12}&\hat{\sigma}_{j11,j21}&0\\
\hat{\sigma}_{j11,j12}&\hat{\sigma}^2_{j12}&0&\hat{\sigma}_{j12,j22}\\
\hat{\sigma}_{j11,j21}&0&\hat{\sigma}^2_{j21}&\hat{\sigma}_{j21,j22}\\
0&\hat{\sigma}_{j12,j22}&\hat{\sigma}_{j21,j22}&\hat{\sigma}^2_{j22}
\end{bmatrix} \quad j=1,2.$$

\noindent With $j,k,l=1,2$, this is a $8\times 8$ matrix, however, given the dependency structure in $\bm{\Sigma}$ resulting from the shared paths between different strategies, $\bm{\hat{\Sigma}}$ maybe singular. Thus, generalized inverse (g-inverse) is used in the definition. 

\begin{theorem}
\label{thm gWald}
Let $T$ be the test statistic defined in Equation (\ref{statistics}). Under the global null hypothesis $H_0: \bm{C\mu}=\bm{0}$, $T$ follows a central $\chi^2$ distribution with degrees of freedom $\nu=rank (\bm{C\Sigma}_0 \bm{C}^T)$, where $\bm{\Sigma}_0$ is the variance-covariance matrix evaluated under the null. Similarly, under a local alternative hypothesis  $H_a: \bm{C\mu}=\bm{\theta}$, for some known constant vector $\bm{\theta}$, it follows a non-central $\chi^2$ distribution with degrees of freedom $ \nu=rank (\bm{C\Sigma}_a \bm{C}^T)$ and the non-centrality parameter $\delta=n\bm{\theta}^T[\bm{C\Sigma}_a \bm{C}^T]^{g} \bm{\theta}$, where $\bm{\Sigma}_a$ is the variance-covariance matrix evaluated under the alternative.
\end{theorem}

Theorem \ref{thm gWald} can be seen as a special case of the Theorem 5.12 of \cite{boos2013essential}, which summarizes the results of \cite{styan1970notes}. We include an outline of the proof in Web appendix A. The test statistic is invariant to the choice of g-inverse and the specification of the contrast matrix $\bm{C}$ as long as $\bm{C}$ is of full row rank. When $\bm{\Sigma}_0$ and $\bm{\Sigma}_a$ are of full rank, the degrees of freedom is equal to $rank(\bm{C})$ and the Wald-type statistic defined in Equation (\ref{statistics}) coincides with the standard Wald statistic defined with the regular inverse. Given the dependency structure in $\bm{\Sigma}$, $T$ with g-inverse is more appropriate in a general SMART setting compared to the standard Wald statistic, and the critical value should be derived from central $\chi^2$ distribution with $rank(\bm{C\Sigma_0C}^T)$ instead of $rank(\bm{C})$. Failing to use the proper degrees of freedom could lead to a higher chance to reject the null hypothesis than nominal level (i.e. overpowered), as observed in \cite{ko2012up}, \cite{ogbagaber2016design}, and \cite{zhong2018design}. 

Even with the proper adjustment of the degrees of freedom for a singular $\bm{\Sigma}$, the distribution of the Wald-type statistic defined in Equation (\ref{statistics}) might differ substantially from its asymptotic distribution, in particular when the sample size $n$ is small. Thus, in addition to adjusting the degrees of freedom of the $\chi^2$ test, we also propose to multiply the estimator of the variance-covariance matrix with a variance inflation factor $\frac{n}{n-p}$, where $p$ is the total number of parameters estimated in a trial. In the example SMART design with 8 strategies in Figure 1, these include $(\kappa_1, p_1, q_1, \pi_j, \mu_{A_jB_k}, \mu_{A_j,C_l}, \sigma_{A_jB_k}, \sigma_{A_jC_l})$ for $j,k,l=1,2$, so the total number of parameters $p=21$.

\section{Interim monitoring in SMART (IM-SMART)}
\label{GS-SMART}

Our goal is to introduce the group sequential methods to SMARTs to allow interim monitoring of trial data. Suppose we have $M$ number of planned analyses in total (i.e., $M-1$ interim analyses). Let $t_m, m=1, \dots,M$ denotes the decision time of the $m^{th}$ analysis, $n_{t_m}$ denotes the cumulative sample size up to time $t_m$, $\bm{\mu}(t_m)$ and $\bm{\Sigma}(t_m)$ denotes the corresponding strategy mean vector and variance-covariance matrix. The test statistic at the $m^{th}$ analysis is
\begin{equation}
    T(t_m)=n_{t_m}\bm{\hat{\mu}}(t_m)^T\bm{C}^T(\bm{C\hat{\Sigma}}(t_m)\bm{C}^T)^g \bm{C\hat{\mu}}(t_m).
\end{equation}

Efficacy boundaries $b_1, \dots, b_M$ are defined such that the trial stops at interim analysis $m^*$ if the global null hypothesis is not rejected at all interim looks $1,\dots,m^*-1$ and is rejected at the $(m^*)^{th}$ look $(T(t_{m^*}) > b_{m^*} )$. Such design that maintains significance level $\alpha$ satisfies the closure testing procedure, that is,
\begin{equation}
\label{closure}
    \sum^M_{m^*=1} Pr(\bigcap^{m^*-1}_{m=1} T(t_m) \le b_m \cap T(t_{m^*}) > b_{m^*} | H_0) = \alpha.
\end{equation}

\noindent If the joint distribution of $T(t_1), \dots, T(t_M)$ is available, given a specific value of $M$, the overall type I error $\alpha$ can be split into $M$ components to calculate different efficacy boundaries such as the Pocock boundary \citep{pocock1977group}, the O'Brien Fleming Boundary \citep{o1979multiple}, or by some pre-specified continuous error spending function \citep{gordon1983discrete}. In traditional group sequential framework, the joint probability in Equation (\ref{closure}) is usually evaluated recursively \citep{armitage1969repeated}. Such calculation requires knowledge about the correlation between test statistics $T(t_m)$ across interim looks. Since $T(t_m)$ is in a quadratic form, we consider the following decomposition. 

\begin{theorem}
\label{thm MN}
Let $\bm{Z}(t_m)$ be a vector of length $\nu$, such that $T(t_m)=\bm{Z}(t_m)^T\bm{Z}(t_m)$ for $m=1,\dots,M$. Under the null hypothesis, treating $\bm{\Sigma}$ as known, the stacked vector $\mathbf{Z}=( \bm{Z}(t_1),\dots, \bm{Z}(t_M) )^T$ follows a multivariate normal distribution
\begin{equation}
    \mathbf{Z} \sim MN_{\nu \times M} (\mathbf{0},\bm{\Lambda} \otimes \bm{I}_\nu),
\end{equation}

\noindent where $\mathbf{0}$ is a vector of zeros, $\bm{\Lambda}$ is a $M \times M$ matrix with diagonal elements $\Lambda_{ii}$=1, and off-diagonal elements $\underset{i <j}{\Lambda_{ij}}=\sqrt{n_{t_i}/n_{t_j}}$, where $n_{t_i}$ is the cumulative sample size upon time $t_i$, $\bm{I}_\nu$ is a $\nu \times \nu$ identity matrix, and $\otimes$ denotes the Kronecker product.
\end{theorem}

Then, $(T(t_1), \dots, T(t_M))^T$ follows a multivariate $\chi^2$ distribution of Wishart type with associated correlation matrix $\bm{\Lambda} \otimes \bm{I}_\nu$ \citep{dickhaus2015survey}. The proof of Theorem \ref{thm MN} is given in Web appendix B.

\subsection{Efficacy boundaries via multivariate $\chi^2$ distribution}
\label{s:boundaries}
Given the information proportion at each time point $m, m=1,\dots,M$, we can use the distribution of $\mathbf{Z}=( \bm{Z}(t_1),\dots, \bm{Z}(t_M) )^T$ outlined in Theorem \ref{thm MN} to  numerically generate the multivariate distribution of the test statistics $(T(t_1), \dots, T(t_M))^T$ under the null hypothesis. For example, suppose the eight ATSs from the SMART in Figure 1 is to be compared in a IM-SMART design that has 2 interim analyses planned at 1/3 (33.3$\%$) and 2/3 (66.7$\%$) information proportions of the total sample size followed by a final analysis at the end of the trial ($M=3$). Suppose under the null hypothesis, each test statistic $T(t_m)$ follows a $\chi^2$ distribution with 5 degrees of freedom. The stacked normalized vector $\mathbf{Z}=(\bm{Z}(t_1), \bm{Z}(t_2), \bm{Z}(t_3))^T$ follows a zero-mean multivariate normal distribution with the variance-covariance matrix 
$$\begin{bmatrix}
1 & \sqrt{1/2} & \sqrt{1/3}\\
\sqrt{1/2} &1 & \sqrt{2/3}\\
\sqrt{1/3}& \sqrt{2/3} &1
\end{bmatrix} \otimes I_5.$$
\noindent To obtain the tri-variate $\chi^2$ distribution, one can numerically generate a sufficiently large sample of $\mathbf{Z}$ and compute the empirical joint distribution of $T(t_m)=\bm{Z}(t_m)^T\bm{Z}(t_m)$ for $m=1,2,3$. To demonstrate, we generated a sample of 100,000 replications from the multivariate normal distribution above using the R package `MASS', and calculated the empirical tri-variate $\chi^2$ distribution. Then, a fine grid search was performed to find the efficacy boundaries satisfying Equation (\ref{closure}) given the overall type I error $\alpha$. When $\alpha=0.05$, for the Pocock-Type boundaries, we have $b_1=b_2=b_3=14.46$, and the O'Brien Fleming-Type boundaries take values $b_1=23.28, b_2=19.00, b_3=13.44$. Other types of boundaries under this setting or with more interim looks can be obtained in a similar manner.   

While straightforward, the above approach for finding boundaries might be computationally intensive in some cases. Thus, it is appealing to have alternative solutions. \cite{dickhaus2015survey} explored methods to approximate the multivariate chi-square distribution with limited dimensions. For example, for $M=2$, the bivariate $\chi^2$ distribution can be approximated by
\begin{equation}
    Pr(T(t_1)<x_1, T(t_2)<x_2;\bm{\Lambda}_{12})=\sum^{\infty}_{n=0}a_n(\lambda_{1}^2, \dots, \lambda_{\nu}^2)G^{(n)}_{\nu/2+n} (\frac{x_1}{2})  G^{(n)}_{\nu/2+n} (\frac{x_2}{2}),
\end{equation}
where $\bm{\Lambda}_{12}=cov(\bm{Z}(t_1), \bm{Z}(t_2))$, cannonical correlations $\lambda_{1}, \dots, \lambda_{\nu}$ are the roots of eigenvalues for $\bm{\Lambda}_{12}\bm{\Lambda}_{21}$, $a_n = a_n(\lambda_{1}^2, \dots, \lambda_{\nu}^2)$ is a set of polynomials defined recursively as
$$
a_0=1, \quad a_{n+1} = \frac{1}{2n+1} \sum^n_{w=0} a_{n-w} \sum^{\nu}_{\mu=1} \lambda_{\mu}^{2(w+1)},
$$

\noindent and $G^{(n)}_{\nu/2+n}(\cdot)$ is the $n^{th}$ derivatives of a Gamma density with parameters $\nu/2+n$ (shape) and 1 (rate). Applying this approximation of bivariate $\chi^2$ distribution to the left-hand side of the closure testing procedure in Equation (\ref{closure}), we can obtain the efficacy boundaries for $M=2$ with different information proportion at interim analysis by a grid search. Table \ref{t:one} shows the boundaries produced by this approximation.

\begin{table}
\caption{Efficacy boundaries for single interim analysis planned at different information proportions and the final analysis, obtained by the approximation method, $M=2$, $\nu=5$, $\alpha=0.05$.}
\label{t:one}
\begin{center}
\begin{tabular}{lllll}
\toprule
Info. prop.   & \multicolumn{2}{l}{Pocock-Type boundaries} & \multicolumn{2}{l}{OBF-Type boundaries} \\ \hline
              & Interim              & Final                & Interim            & Final              \\ \cline{2-5} 
0.20          & 12.72                & 12.72                & 24.78              & 11.08              \\
0.30          & 12.66                & 12.66                & 20.28              & 11.11              \\
0.40          & 12.59                & 12.59                & 17.68              & 11.18              \\
\textbf{0.50} & \textbf{12.50}       & \textbf{12.50}       & \textbf{15.94}     & \textbf{11.27}     \\
0.60          & 12.39                & 12.39                & 14.68              & 11.37              \\
0.70          & 12.26                & 12.26                & 13.72              & 11.48              \\
0.80          & 12.08                & 12.08                & 12.91              & 11.55              \\
0.90          & 11.85                & 11.85                & 12.17              & 11.55              \\ \bottomrule
\end{tabular}
\end{center}
\end{table}

\subsection{Identifying the best ATS(s)}
\label{post hoc MC}
Rejection of the global null hypothesis at either an interim or the final analysis in a IM-SMART design implies that there is at least one ATS that is significantly different from another ATS. One might be interested in identifying the most effective ATS(s) in the trial. Taking into account of the correlation between treatment estimates and multiplicity, we employ the method proposed by \cite{ertefaie2016identifying} to perform a \textit{post hoc} test to identify the best strategies whenever the global null is rejected. Suppose among all eight strategies in the SMART setting aforementioned, there is a true set of best ATSs, $\mathbb{S}$, with the maximum (or the best) outcome, and let $\hat{\mathbb{S}}_{m}$ denote a set of ATSs that are not statistically different from $\mathbb{S}$ based on data accumulated by interim look at time $t_m, m=1, \dots, M$. Thus, for a $M$-look group sequential SMART design, the overall probability to identify the best ATSs is
\begin{equation}
\label{bestselect}
    best.select=\sum^M_{m^*=1} Pr(  \mathbb{S} \in \hat{S}_{m^*} \; \big{|} \; \bigcap^{m^*-1}_{m=1} T^g(t_m) \le b_m \cap T^g(t_{m^*})>b_{m^*}),
\end{equation}

\noindent where $ \hat{S}_{m^*}$ is estimated by Ertefaie's resampling method at each interim look whenever the global null hypothesis is rejected.

\section{Simulation}
\label{s:simulation}

To evaluate the operating characteristics of the proposed IM-SMART design, we conducted a variety of simulations. The simulation scenarios were designed based on the SMART described in Figure 1. Under each scenario, we first randomly assigned each patient to the initial treatment $A_j$ with probability $\kappa_j^{-1}$ for $j=1,2$. The response $R_i$ for patient $i$ was generated from a Bernoulli distribution with probability of success $\pi_j$ if the patient was initially assigned to treatment $A_j, j=1,2$. Given the initial assignment and the response, we then simulated the second stage treatment assignment with probability $p_1$ for treatment $B_1$ among responders, and with probability $q_1$ for treatment $C_1$ among non-responders. After obtaining the realization of treatment paths for each patient in the simulated dataset, we generated the counterfactual outcome variables $Y(A_jB_k)$ and $Y(A_jC_l)$ from normal distributions with means $\mu_{A_jB_k}$ or $\mu_{A_jC_l}$, and standard deviations $\sigma_{A_jB_k}$ and $\sigma_{A_jC_l}$, respectively for $j,k,l=1,2$. The observed outcome $Y_i$ for the $i^{th}$ patient in the sample is then taken as the counterfactual outcome consistent with the patient's treatment path.

We followed \cite{ko2012up} and \cite{ogbagaber2016design} for specification of the simulation parameters. Specifically, under the null hypothesis, we set $\mu_{A_jB_k}=\mu_{A_jC_l}=15$, $\forall j,k,l=1,2$; while under the alternative hypothesis, we set $\mu_{A_jB_1}=\mu_{A_jC_2}=15$, $\mu_{A_jB_2}=22$ and $\mu_{A_jC_1}=20$, $\forall j=1,2$. For the true values of variances, we set $\sigma_{A_jB_k}=12$ and $\sigma_{A_jC_l}=10$, $\forall j,k,l=1,2$. The four alternative scenarios we explored are established by varying the first stage response rate $\pi_j$ ($(\pi_1,\pi_2)=\{ (0.5,0.5), (0.2,0.5), (0.7,0.5), (0.2,0.7)\}$). The strategy means of the four different alternative scenarios can be calculated by plugging-in these settings to Equation (2) and they are shown in Table \ref{t:scenarios}. For example, under alternative scenario 1, the strategy mean for ATS $A_1B_1C_1$, $\mu_{111}=\pi_1*\mu_{A_1B_1}+(1-\pi_1)*\mu_{A_1C_1}=0.5*15+0.5*20=17.5$. We also varied the second stage randomization probability $p_1$ for responders ($p_1 = \{0.5,0.7,0.8 \}$) under each of the scenarios . For an easier comparison across scenarios, we assumed equal randomization for initial treatments (i.e. $\kappa_1=\kappa_2=2$) and second stage treatment for non-responders (i.e. $q_1=q_2=0.5$). 

\begin{table}
\caption{Strategy means under alternative hypothesis for different scenarios in the simulation study, calculated by plugging-in the simulation setups to Equation (2).}
\label{t:scenarios}
\begin{center}
\begin{tabular}{lllllllll}
\toprule
           & $\mu_{111}$ & $\mu_{112}$ & $\mu_{121}$ & $\mu_{122}$ & $\mu_{211}$ & $\mu_{212}$ & $\mu_{221}$ & $\mu_{222}$ \\ \hline
Alt(1) & 17.5                        & 15.0                        & 21.0                        & 18.0                        & 17.5                        & 15.0                        & 21.0                        & 18.5                        \\
Alt(2) & 19.0                        & 15.0                        & 20.4                        & 16.4                        & 17.5                        & 15.0                        & 21.0                        & 18.5                        \\
Alt(3) & 16.5                        & 15.0                        & 21.4                        & 19.0                        & 17.5                        & 15.0                        & 21.0                        & 18.5                        \\
Alt(4) & 19.0                        & 15.0                        & 20.4                        & 16.4                        & 16.5                        & 15.0                        & 21.4                        & 19.9                        \\ \bottomrule
\end{tabular}
\end{center}
\end{table}

We planned single interim analysis when the information proportion is 0.5, which means the interim analysis is triggered at half of the maximum sample size. However, these results can be extended to other information proportions and to more than one interim analysis (see Section 3.2). Using Equations (\ref{IPWN}), (\ref{var}) and (\ref{cov}), we estimated the strategy means and the variance-covariance matrix for each sample. We would like to note here that under the null hypothesis, $\bm{\Sigma}_0$ is singular and has rank 6 for all scenarios by plugging in the true parameters. Thus, with a contrast matrix $\bm{C}= [ \mathbb{1}_7, -1*\bm{I}_{7} ]$, the global Wald test statistic $T$ follows a $\chi^2$ distribution with 5 degrees of freedom. For interim and final decisions, we used the efficacy boundaries shown in Table \ref{t:one} at $0.5$ information proportion. Under each scenario, we simulated 5000 trials and evaluated the empirical type I error (proportion of samples for which the null hypothesis is rejected) and the power (rejection rate under the alternative scenarios). 

The results under various null scenarios are shown in Table \ref{t:three}. For each scenario, we evaluated the overall type I error rate under two sample sizes: $n=n_{max}$, which is the estimated maximum sample size to achieve $90\%$ power at the final analysis, and a larger sample size $n=500$, to assess the asymptotic behavior of the test statistic. The type I error is generally well-controlled around the nominal rate of $\alpha=0.05$ for almost all null scenarios at $n=n_{max}$. For the larger sample size, the empirical type I error closely approximates the nominal level.

\begin{table}
\caption{False rejection rates for the proposed IM-SMART design under null scenarios with interim analysis taking place at information proportion of $50\%$. Different null scenarios are defined by varying the response rate for initial treatments $A_j$, $\pi_j$, for $j=1,2$, and the probability to re-randomize responders to treatment $B_1$, $p_1$, in the second stage. The empirical type I error is evaluated under two sample size for the final analysis, $n= \{ n_{max}, 500 \}$, and over 5000 simulated trials.}
\label{t:three}
\begin{center}
\begin{threeparttable}
\begin{tabular}{llllllllll}
\toprule
                          &         &       &           & \multicolumn{3}{c}{$n=n_{max}$}        & \multicolumn{3}{c}{n=500}              \\ \hline
                          $\pi_1$ & $\pi_2$ & $p_1$ & $n_{max}$ & Rej at & Rej at & Empirical          & Rej at & Rej at & Empirical          \\
                          &         &       &           & Interim   & Final     & Type I error          & Interim   & Final     & Type I error          \\ \hline
                          \multicolumn{10}{c}{Pocock-Type Boundary}                                                                               \\ \hline
 0.5     & 0.5     & 0.5   & 252       & 0.021     & 0.025     & \textbf{0.045} & 0.024     & 0.022     & \textbf{0.045} \\
                           0.5     & 0.5     & 0.8   & 317       & 0.039     & 0.025     & \textbf{0.063} & 0.035     & 0.024     & \textbf{0.058} \\
 0.2     & 0.5     & 0.5   & 263       & 0.024     & 0.023     & \textbf{0.046} & 0.025     & 0.024     & \textbf{0.048} \\
                           0.2     & 0.5     & 0.8   & 309       & 0.031     & 0.021     & \textbf{0.051} & 0.028     & 0.023     & \textbf{0.050} \\
 0.7     & 0.5     & 0.5   & 244       & 0.023     & 0.017     & \textbf{0.040} & 0.028     & 0.022     & \textbf{0.049} \\
                           0.7     & 0.5     & 0.7   & 273       & 0.029     & 0.021     & \textbf{0.049} & 0.023     & 0.026     & \textbf{0.048} \\
 0.2     & 0.7     & 0.5   & 254       & 0.019     & 0.022     & \textbf{0.041} & 0.025     & 0.025     & \textbf{0.049} \\
                           0.2     & 0.7     & 0.7   & 278       & 0.031     & 0.023    & \textbf{0.053} & 0.024     & 0.019     & \textbf{0.043} \\ \hline
                          \multicolumn{10}{c}{OBF-Type Boundary}                                                                                  \\ \hline
 0.5     & 0.5     & 0.5   & 228       & 0.007     & 0.040     & \textbf{0.046} & 0.007     & 0.044     & \textbf{0.051} \\
                           0.5     & 0.5     & 0.8   & 288       & 0.010     & 0.052     & \textbf{0.062} & 0.009     & 0.044     & \textbf{0.053} \\
 0.2     & 0.5     & 0.5   & 239       & 0.003     & 0.039     & \textbf{0.041} & 0.005     & 0.042     & \textbf{0.047} \\
                           0.2     & 0.5     & 0.8   & 280       & 0.009     & 0.044     & \textbf{0.052} & 0.007     & 0.047     & \textbf{0.054} \\
 0.7     & 0.5     & 0.5   & 221       & 0.005     & 0.035     & \textbf{0.040} & 0.007     & 0.050     & \textbf{0.057} \\
                           0.7     & 0.5     & 0.7   & 248       & 0.008     & 0.041     & \textbf{0.048} & 0.009     & 0.042     & \textbf{0.050} \\
 0.2     & 0.7     & 0.5   & 230       & 0.003     & 0.036     & \textbf{0.039} & 0.008     & 0.040     & \textbf{0.047} \\
                           0.2     & 0.7     & 0.7   & 252       & 0.007     & 0.039     & \textbf{0.045} & 0.008     & 0.043     & \textbf{0.051} \\ \bottomrule
\end{tabular}
\begin{tablenotes}
\item Notes: Rej at Interim: probability of rejection at interim analysis; Rej at Final: probability of rejection at final analysis condition on do not reject at interim analysis; $n_{max}$: the maximum sample size required to achieve $90\%$ power at final analysis.
\end{tablenotes}
\end{threeparttable}
\end{center}
\end{table}

Table \ref{t:four} summarizes the performance of the group sequential analysis with sample size $n_{max}$ under the alternative scenarios. In addition to the empirical rejection rates at the interim, the conditional rejection rate at the final analysis (if the global null hypothesis is not rejected at the interim), and the empirical power, Table \ref{t:four} also presents the average sample size at which the trial stops ($E(n)$) and the sample size required to achieve $90\%$ power in the absence of an interim analysis ($n_{classical}$) according to \cite{ogbagaber2016design}. For the Pocock-type boundaries, under all different scenarios, the rejection rate at interim is $50\%$ or lower. The overall empirical power is close to the designed level of $90\%$. The estimated expected sample sizes (the sample size at which the trial stops) for all scenarios are considerably lower than that of a classical SMART design with no interim analysis. In general, IM-SMART design saves about $15\%-20\%$ of the total sample size on average. The performance of the OBF-type boundaries shows a similar trend. Since OBF boundaries are more conservative at the interim analysis, the rate of early stopping is less than it would be when using Pocock-type boundaries. The estimated expected sample size is larger than that of the Pocock-type boundaries, but it still leads to about a $10\%$ reduction in expected total sample size on average.

\begin{table}
\caption{Power of the proposed IM-SMART design under alternative scenarios with interim analysis taking place at information proportion of $50\%$. Different alternative scenarios are defined in Table \ref{t:scenarios} and varying the response rate for initial treatments $A_j$, $\pi_j$, for $j=1,2$, and the probability to re-randomize responders to treatment $B_1$, $p_1$, in the second stage. The empirical power is evaluated by the percentage of any rejection of the global null hypothesis either at the interim or final analysis, and Best.select is the probability of correctly selecting the best ATSs. 5000 trials are simulated under each scenarios.}
\label{t:four}
\begin{center}
\begin{threeparttable}
\begin{tabular}{lllllllllll}
\toprule
                          & $\pi_1$ & $\pi_2$ & $p_1$ & Rej at & Rej at & Empirical & Best. & $E(n)$         & $n_{classical}$ & $n_{max}$ \\
                          &         &               &           & Interim   & Final     & Power     &select             &                 &    &  \\ \hline
                          & \multicolumn{10}{c}{Pocock-Type Boundary}                                                                       \\ \hline
\multirow{2}{*}{Alt(1)} & 0.5     & 0.5     & 0.5    & 0.43     & 0.78    & 0.88     & 0.86       & \textbf{198} & 225 & 252  \\
                          & 0.5     & 0.5     & 0.8     & 0.52     & 0.79     & 0.90     & 0.87  & \textbf{235} & 283 & 317 \\
\multirow{2}{*}{Alt(2)} & 0.2     & 0.5     & 0.5    & 0.45     & 0.78     & 0.88     & 0.88       & \textbf{204} & 235 & 263 \\
                          & 0.2     & 0.5     & 0.8   & 0.50     & 0.80     & 0.90     & 0.89     & \textbf{232} & 276 & 309  \\
\multirow{2}{*}{Alt(3)} & 0.7     & 0.5     & 0.5   & 0.43     & 0.78     & 0.87     & 0.87       & \textbf{192} & 218 & 244  \\
                          & 0.7     & 0.5     & 0.7   & 0.47     & 0.79     & 0.89     & 0.88     & \textbf{208} & 244 & 273   \\
\multirow{2}{*}{Alt(4)} & 0.2     & 0.7     & 0.5  & 0.43     & 0.79     & 0.88     & 0.88       & \textbf{199} & 226 & 254   \\
                          & 0.2     & 0.7     & 0.7  & 0.47     & 0.79     & 0.89     & 0.88  & \textbf{213} & 248  & 278   \\ \hline
                          & \multicolumn{10}{c}{OBF-Type Boundary}                                                                          \\ \hline
\multirow{2}{*}{Alt(1)} & 0.5     & 0.5     & 0.5    & 0.22     & 0.84     & 0.88     & 0.86  & \textbf{203} & 225 & 228\\
                          & 0.5     & 0.5     & 0.8    & 0.28     & 0.83     & 0.88     & 0.86     & \textbf{248} & 283 & 288  \\
\multirow{2}{*}{Alt(2)} & 0.2     & 0.5     & 0.5    & 0.21     & 0.84     & 0.87     & 0.87     & \textbf{214} & 235 & 239 \\
                          & 0.2     & 0.5     & 0.8    & 0.31     & 0.88     & 0.92     & 0.92    & \textbf{237} & 276 & 280 \\
\multirow{2}{*}{Alt(3)} & 0.7     & 0.5     & 0.5   & 0.20     & 0.83     & 0.86     & 0.86   & \textbf{199} & 218  & 221    \\
                          & 0.7     & 0.5     & 0.7   & 0.23     & 0.82     & 0.87     & 0.86   & \textbf{220} & 244  & 248  \\
\multirow{2}{*}{Alt(4)} & 0.2     & 0.7     & 0.5   & 0.22     & 0.84     & 0.88     & 0.87     & \textbf{205} & 226  & 230   \\
                          & 0.2     & 0.7     & 0.7    & 0.23     & 0.83     & 0.87     & 0.87   & \textbf{223} & 248  & 252\\ \bottomrule
\end{tabular}
\begin{tablenotes}
\item Notes: Rej at Interim: probability of rejection at interim analysis; Rej at Final: probability of rejection at final analysis condition on do not reject at interim analysis; $E(n)$: average sample size at which the trial stops; $n_{classical}$: sample size required to achieve $90\%$ power for classical SMART design; $n_{max}$: the maximum sample size required to achieve $90\%$ power at final analysis adjusted for interim look.
\end{tablenotes}
\end{threeparttable}
\end{center}
\end{table}

The $Best.select$ column in Table \ref{t:four} demonstrates the probability of correctly selecting the best treatment strategies (with the maximum outcome) under each alternative scenarios via the procedure described in Section \ref{post hoc MC}. The probability is very close to the empirical power, implying that the IM-SMART design is powerful in identifying the best ATSs.

\section{Analysis of RAPID study}
\label{s:rapid}

In this Section, we re-analyze the RAPID (Reducing pain; Preventing depression) SMART trial \citep{karp2019improving} using IM-SMART approach. The goal of this trial was to evaluate the effects of cognitive behavioral therapy (CBT) and physical therapy (PT) in patients with knee osteoarthritis and comorbid subsyndromal depressive symptoms. It is worth mentioning that any interim analysis should be planned \textit{a priori}, and this application is only for the purpose of demonstration. Moreover, this study is relatively small-sized compared to the sample sizes considered in Section \ref{s:simulation}, and it was not designed for the global Wald test we apply here. Thus, the test might be under-powered. Nevertheless, this application signifies the benefit of adding interim analyses to SMARTs. 

In the RAPID study, 99 patients were randomized to CBT (n=39), PT (n=40), and enhanced usual care (EUC, n=20). After 8 weeks of the initial treatment, patients' responses were assessed by improvement in the P-GIC (patient-reported global impression of change) score in activity limitations, symptoms, emotions, and overall quality of life. A result of at least `much better' for the P-GIC score is defined as a response. Non-responders were re-randomized to either having an additional 4 weeks of their initial treatment or switching to the other non-EUC treatment. 

We will analyze the RAPID data in two different ways to demonstrate how our method could be applied to various SMART settings. The first analysis will exclude the EUC arm, so that the design becomes very similar to the one considered in the method section. The second analysis will analyze the data including the EUC arm. In the first case, there are 4 ATSs in this SMART, namely: 

\begin{enumerate}
    \item Start on 8 weeks of CBT, if no response, continue CBT for 4 more weeks (CBT-CBT);
    \item Start on 8 weeks of CBT, if no response, switch to PT for 8 weeks (CBT-PT);
    \item Start on 8 weeks of PT, if no response, switch to CBT for 8 weeks CBT (PT-CBT);
    \item Start on 8 weeks of PT, if no response, continue PT for 4 more weeks (PT-PT).
\end{enumerate}

All patients were followed for 12 months regardless of their initial treatment assignment or response status. We considered three of the continuous outcomes measured at the last follow-up visit in this application: a measure of pain (PAIN, numeric rating score), depression (using PHQ-9), and anxiety (using GAD-7) \citep{karp2019improving}. Excluding patients whose outcomes of interest are missing, we have 65 patients in the analysis set (Figure 2). 

\begin{figure}
 \centerline{\includegraphics[width=5.25in]{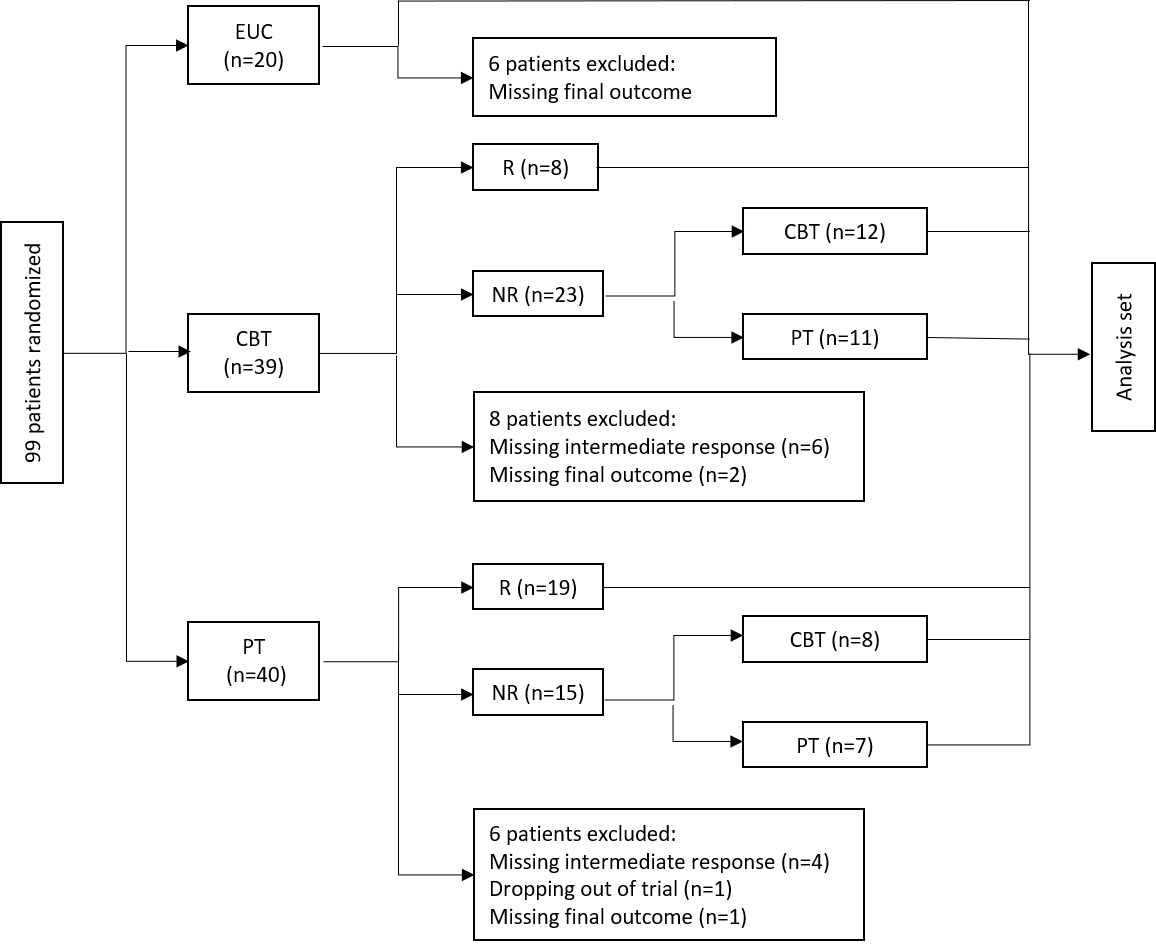}}
\caption{Patient flow for RAPID study. There are three treatment options, EUC (Enharced Usual Care), CBT (Cognitive Behavioral Therapy), and PT (Physical Therapy). The intermediate responders (R) and non-responders (NR) are determined by patients' P-GIC (patient-reported global impression of change in activity limitations, symptoms, emotions and overall quality of life) score.}
\end{figure}

The question of interest is whether these three aforementioned outcome measures differ across these four ATSs, and whether the differences, if any, can be detected early with IM-SMART design. For each of the outcomes, the global null hypothesis is that all of the four ATSs have the same strategy means. Under the null hypothesis, the global Wald-type test statistic follows a $\chi^2$ distribution with 3 degrees of freedom. To mimic the process of conducting an interim analysis planned \textit{a priori}, we assumed the enrollment of patients and the evaluation of outcomes follow the order of their study ID. Suppose the interim analysis is conducted after 46 patients were evaluated for their outcomes (i.e. at $70\%$ of total information), the interim  and final critical values according to the Pocock-type boundaries are both 8.83, which is obtained following the steps described in Section \ref{s:boundaries}. Table \ref{t:rapid} shows results for the interim and final analyses for each of the outcomes of interest.

\begin{sidewaystable}
\caption[Re-analysis of RAPID data.]%
{Re-analysis of RAPID data, where a interim analysis is assumed to take place at $70 \%$ of total information. The strategy means and variances are estimated by Equation (\ref{IPWN}) and (\ref{var}). The decisions are made by comparing the test statistic $\chi^2_{d.f.=3}$ with a Pocock-type critical value $b_1=b_2=8.83$ for each of the outcomes of interest.}
\label{t:rapid}
\begin{center}
\begin{tabular}{lllllllll}
\toprule
 & PHQ-9 &  &  & GAD-7 &  &  & PAIN &  \\ \cline{2-3} \cline{5-6} \cline{8-9} 
 & Interim & Final &  & Interim & Final &  & Interim & Final \\ \cline{2-3} \cline{5-6} \cline{8-9} 
 Strategies & Mean (Var) & Mean (Var) &  & Mean (Var) & Mean (Var) &  & Mean (Var) & Mean (Var) \\ \hline
CBT-CBT & 2.61 (0.37) & 2.62 (0.50) &  & 1.59 (0.34) & 1.63 (0.29) &  & 4.77 (0.73) & 5.42 (0.69)  \\
CBT-PT & 3.43 (1.44) & 3.23 (0.79) &  & 3.52 (0.73) & 3.28 (0.42) &  & 5.90 (2.17) & 6.24 (1.19)  \\
PT-CBT & 4.01 (0.67) & 3.43 (0.35) &  & 2.53 (0.47) & 2.45 (0.28) &  & 5.54 (1.21) & 5.38 (0.85) \\
PT-PT & 4.42 (0.85) & 3.64 (0.54) &  & 1.21 (0.20) & 1.30 (0.12) &  & 6.04 (1.09) & 5.27 (0.75) \\
$\chi^2_{d.f.=3}$ & 2.804 & 1.039 &  & 8.955 & 11.090 &  & 1.121 & 0.551\\
Decision & Continue & Do not reject& & Stop & Reject & & Continue & Do not reject \\
\bottomrule
\end{tabular}
\end{center}
\end{sidewaystable}

At the interim analysis, the test statistics for outcome PHQ-9 is below the critical value, so patient enrollment would have continued beyond the interim if PHQ-9 was the primary outcome. The final analysis based on the whole sample agrees with the interim result that there is no significant difference between treatment strategies based on the PHQ-9 measure. The results for GAD-7 is in the other direction. The interim test statistic for GAD-7 is larger than the critical value, implying an early stop for efficacy if GAD-7 was our primary outcome. The final test statistic agrees with the interim decision. Had interim analysis been included in the trial design, it would have lead to a maximum of $30\%$ reduction in the total sample size. The PAIN score follows similar pattern as the PHQ-9 score. 

For GAD-7 score, since the global null hypothesis is rejected at the interim and final analysis, a \textit{post hoc} analysis is triggered. The goal of this \textit{post hoc} analysis is to find the best ATSs (i.e. with lowest GAD-7) among the four. Using the multiple comparison procedure proposed by \cite{ertefaie2016identifying}, this is equivalent to identify strategies with index $i$ that $\hat{\mu}_i \le min(\hat{\mu}_j-c_i\hat{\sigma}_{ij})$ for all $j \ne i$, where $\hat{\mu}_i$ is the strategy mean with index $i$, $\hat{\sigma}_{ij}$ is the pooled variance between the two strategies that are being compared, and $c_i$ is a constant, following Ertafie's resampling method, can be estimated to account for the correlation between treatment strategies while maintaining overall type I error. For the interim readouts, the estimated values are $c_{CBT-CBT}=1.908$, $c_{CBT-PT}=1.598$, $c_{PT-CBT}=1.726$, $c_{PT-PT}=1.766$, and strategies CBT-CBT and PT-PT are identified as the best ATSs. For the final analysis, $c_{CBT-CBT}=1.674$, $c_{CBT-PT}=1.715$, $c_{PT-CBT}= 1.679$, $c_{PT-PT}=1.826$. The best ATSs selected are the same as the interim analysis.

In the second analysis, we include the EUC as a stand-alone strategy, and altogether we will compare 5 strategies. Since patients on the EUC arm were not examined for the intermediate endpoint and were not re-randomized during the study, we estimated the mean treatment effect and variance for EUC arm by the observed sample mean and variance. Under the null hypothesis, the global Wald-type test statistics follows a $\chi^2$ distribution with 4 degrees of freedom. Similar to the first analysis, a hypothetical interim look is planned at $70\%$ of total information, and the interim and final critical values (Pocock-type) of 10.59 are used. The results are similar to the first analysis (Web table 1).

Similar to the first analysis, a \textit{post hoc} analysis would be triggered for GAD-7 score. Since a control arm (EUC) is included in this study, the goal of this \textit{post hoc} analysis is to test whether any of the four ATSs performs better than EUC. A strategy has significantly lower GAD-7 compared to EUC when $(\hat{\mu}_{EUC}-\hat{\mu}_{i})/\hat{\sigma}_{i} \ge c$, where $\hat{\mu}_{i}$ is the estimated strategy effect for each ATSs other than EUC, $i$ is an arbitrary index, $\hat{\mu}_{EUC}$ is the estimated treatment effect of the EUC arm, $\hat{\sigma}_{i}$ is the pooled standard error between EUC and the strategy indexed with $i$, In this analysis, $c$ takes values 1.714 and 1.546 for the interim and final analyses, respectively. At both the interim and final analysis, strategies CBT-CBT and PT-PT show a significantly lower GAD-7 score compared to the EUC arm. Overall, this application highlights the favorable resource-saving benefit of IM-SMART design, and the flexibility of the proposed method when extended to various settings.

\section{Discussion}
\label{s:discuss}
SMART is an efficient way to evaluate multiple ATSs, but it is generally more resource intensive than traditional designs. We introduced interim monitoring to SMARTs and used simulations to show that the proposed IM-SMART design can maintain overall power and type I error while reducing the expected sample size. This resource-saving benefit of IM-SMART was also illustrated by re-analyzing the RAPID data.

The key feature of our proposed IM-SMART design is that it allows early stopping at the interim analysis when the early efficacy signal is sufficiently strong. In the traditional group sequential framework, it is also common to include futility boundaries at interim analyses, where one can claim early stop if there is strong evidence in favor of the null hypothesis or the direction is opposite to the alternative. From an ethical perspective, it would be less beneficial to enroll more patients to the inferior treatment arms or, sometimes, to continue the trial. The IM-SMART design only considers efficacy boundaries for the global null hypothesis because SMARTs are usually designed to compare strategies combining treatments that have been approved individually for safety. Thus, a futility analysis for safety may not be warranted in this setting. 

In this paper, we utilized two commonly used boundaries, the Pocock and O'Brian Fleming-types, to demonstrate our idea of group sequential testing in the SMART setting. The Pocock-type has constant boundaries across interim looks, while the O'Brian Fleming-type is more stringent in early stopping. Both types of boundaries are commonly used, and the simulation results are consistent with the trend described in traditional group sequential designs. The proposed method can also be extended to other types of boundaries that are discussed in details in \cite{jennison1999group}.

Our proposed method is developed based on a global test between all treatment strategies and the null hypothesis is written as $H_0: \bm{C\mu}=\bm{0}.$ In this form, it is assumed that we are interested in a linear contrast. In some cases, researchers may also want to investigate some non-linear contrast between treatment strategies, where the null hypothesis can be written as $H_0: \bm{h(\mu)}=\bm{0}.$ As discussed in \cite{boos2013essential}, the test statistic follows a similar asymptotic distribution under straightforward application of the Delta method. However, earlier studies also point out that, with finite sample size, the numerical outcomes of such test may be substantially different for different forms of the non-linear function $\bm{h}$ that are algebraically equivalent under the same null hypothesis \citep{gregory1985formulating, lafontaine1986obtaining}. Moreover, researchers may be more interested to claim early stop only if there is an ATS that is evidently superior to all the others or to a common control, rather than the general global test considered in this paper. We proposed to perform pair-wise comparisons as  \textit{post hoc} test. However, more direct answer, for example, restricted inference on 'one-sided' alternative hypothesis, may be of interest. In addition, if researchers are interested in deeply tailored ATSs as secondary aim of the study, the aim could be compromised by early stopping. Future work is needed to investigate how current method can be extended to accommodate these considerations.  

An important practical limitation of such group sequential analysis is the information and time gap between the proposed number of patients to be enrolled before the interim analysis and the time when their outcomes are available for evaluation. Continuing enrollment during this gap is sometimes known as trial `overrun' \citep{whitehead1992overrunning}, if the interim decision is early stopping. The challenge of overrun is inherited from the traditional group sequential design, and it might be problematic when time-to-event outcomes that take a long time to observe are considered in the trial. Since follow-up is generally longer in SMARTs to accommodate multiple stages of randomization and follow-up, this might post a practical challenging during implementation. In our simulation study and in the re-analysis of the RAPID study, we assumed that enrollment is suspended until the responses are evaluated in all patients enrolled before the interim analysis. In practice, this may not be feasible because of the timeline and budget constraints. There are several statistical procedures proposed in literature to account for overrun data in traditional group sequential framework, including Whitehead’s (1992) deletion method and the combination test approach \citep{hall2008sequential,hampson2013group}. Incorporating this consideration in estimating strategy effects at the final analysis and identifying the best timing for interim analysis in the SMART setting can be explored in future research.

Our proposed method is based on a specific type of a two-stage SMART design. However, it can be easily adapted to the other SMART designs, for example where only non-responders will be re-randomized at the second stage, as shown in our application to the RAPID study. Moreover, the method is developed based on a continuous outcome. Future research can generalize this IM-SMART design to include other types of outcomes such as binary or survival endpoints. The approximation of efficacy boundaries for more than 2 interim looks is also of interest, where the challenge is to obtain a closed-form boundary from a multivariate $\chi^2$ distribution.


\section*{Acknowledgements}
We thank the Center for Research Computing at the University of Pittsburgh for providing computational support. We thank Dr. Jordan Karp for providing the RAPID study data.

\bibliographystyle{unsrtnat}
\bibliography{myreference}  

\section*{Supporting information}
Web Appendices referenced in Sections 2, 3 and 5 are available with this paper at the Biometrics website on Wiley Online Library. Source code is available at https://github.com/LWu-pitt/IM-SMART.

\end{document}